\documentclass{emulateapj}
\newcommand{\kms}{km s$^{-1}\;$}

\newcommand{\vmax}{$V_{max}\;$}
\newcommand{\msun}{M$_\odot$}

\shorttitle{Star Formation in Local Volume Galaxies}
\shortauthors{J. C. Lee et al.}

\begin{document}

\title{The Star Formation Demographics of Galaxies in the Local Volume} 

\author{Janice C. Lee\altaffilmark{1,2}, 
Robert C. Kennicutt\altaffilmark{3,4}, 
Jose G. Funes, S.J.\altaffilmark{5}, 
Shoko Sakai\altaffilmark{6}, 
Sanae Akiyama\altaffilmark{4}}

\altaffiltext{1}{Carnegie Observatories, 813 Santa Barbara Street, Pasadena, CA, 91101; jlee@ociw.edu}
\altaffiltext{2}{Hubble Fellow}
\altaffiltext{3}{Institute of Astronomy, University of Cambridge, Madingley Road, Cambridge, CB3 0HA, UK} 
\altaffiltext{4}{Steward Observatory, University of Arizona, 933 N. Cherry Ave., Tucson, AZ 85721}
\altaffiltext{5}{Specola Vaticana, V-00120, Citta del Vaticano, ITALY} 
\altaffiltext{6}{Division of Astronomy and Astrophysics, University of California, Los Angeles, CA 90095}

\begin{abstract}
We examine the connections between the current global star formation 
activity, luminosity, dynamical mass and morphology of galaxies in the Local Volume,
using H$\alpha$ data from the 11 Mpc H$\alpha$ and 
Ultraviolet Galaxy Survey (11HUGS).  
Taking the equivalent width (EW) of the H$\alpha$ emission line as a
tracer of the specific star formation rate, we
analyze the distribution of galaxies in the $M_B$-EW and rotational 
velocity ($V_{max}$)-EW planes.  Star-forming galaxies 
show two characteristic transitions in these planes.  A narrowing of the galaxy locus occurs at $M_B\sim-15$ and $V_{max}\sim$ 50 km s$^{-1}$, where the scatter in the logarithmic EWs drops by a factor of two as the luminosities/masses increase, and galaxy morphologies shift from predominately irregular to late-type spiral. Another transition occurs at $M_B\sim-19$ and $V_{max}\sim$120 km s$^{-1}$, above which the sequence turns off toward lower EWs and becomes mostly populated by intermediate and early-type bulge-prominent spirals.  Between these two transitions, the mean logarithmic EW appears to remain constant at 30\AA.  We comment on how these features reflect established empirical relationships, and provide clues for identifying the large-scale physical processes that both drive and regulate star formation, with emphasis on the low-mass galaxies that dominate our approximately volume-limited sample.
\end{abstract}

\section{Introduction}

It is well established 
that many of the global properties of present-day galaxies scale 
coherently with one another.  Of particular interest is the star 
formation activity; understanding and deciphering trends with 
this property provides insight into the astrophysics that control 
the pace and extent of a galaxy's evolution.  Perhaps the most 
well-known trend is that the current star formation rate (SFR)
varies monotonically along the Hubble sequence (Roberts 1963; Kennicutt 1998).  The scale begins at the ellipticals, which generally have red colors, little gas and essentially no star formation, and proceeds through the lenticulars, spirals and irregulars, which progressively have bluer colors, larger gas fractions and greater relative levels of star formation.

The quantification of these patterns began with the development of SFR measurement diagnostics (e.g., Tinsley 1968; Kennicutt 1983a), and is now reaching full maturity with the application of such techniques to large, statistically complete samples of local galaxies (e.g., Brinchmann et al. 2004, Salim et al. 2007), as well as of galaxies at higher redshift (e.g. Noekse et al. 2007a).  Progress has been rapid in quantifying relations for galaxies in the intermediate and high-end of the luminosity function, with the significant result that the galaxy distribution is bimodal and may divide into distinct populations of old, red spheroids and blue star forming disks at a stellar mass of $\sim3\times10^{10}$M$_{\odot}$ (Kauffmann et al. 2003).

In contrast, progress on the dwarf galaxy population has been slower, primarily due to the lack of complete and statistically robust samples in the low-mass regime.  To address this need, we have focused on our nearest neighbors and are taking an appoximately volume-limited (and hence dwarf-dominated) inventory of star formation in the Local Volume.  Our 11 Mpc H$\alpha$ and UV Galaxy Survey (11HUGS) has obtained narrowband H$\alpha$ emission-line imaging, which traces massive O-star formation, and is also collecting GALEX UV imaging, which traces the photospheric emission of O-type as well as longer-lived B-stars.  11HUGS also serves as the basis for the Spitzer Local Volume Legacy program, an IRAC (mid-IR) and MIPS (far-IR) imaging survey of 258 galaxies, which will capture the energy which is absorbed and re-radiated by dust.
The 11HUGS parent sample contains $\sim$400 galaxies and includes all known spirals, irregulars, as well as star-forming early-type galaxies with $d < $ 11 Mpc, $|b| >$ 20 \degr and $B <$ 15 mag.  This Letter takes a first look at results from the recently completed H$\alpha$ imaging component of the survey (Lee 2006, Kennicutt et al. 2008), and explores trends in the magnitude vs. H$\alpha$ equivalent width
(EW) plane, a close cousin of the more commonly studied color-magnitude diagram (e.g. Baldry et al. 2004 and references therein). 

\begin{figure*}[t]
\epsscale{1.05}
\plottwo{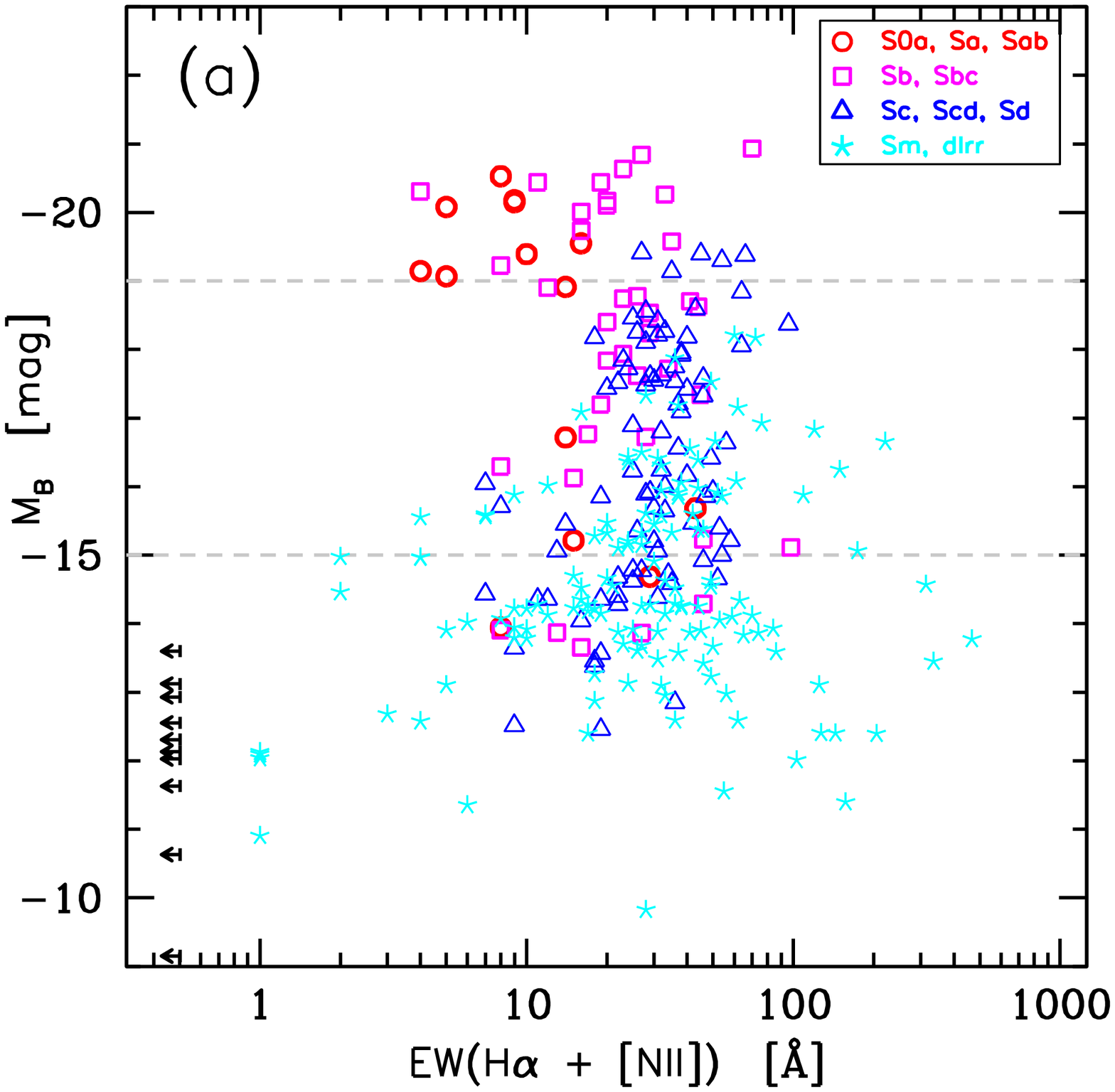}{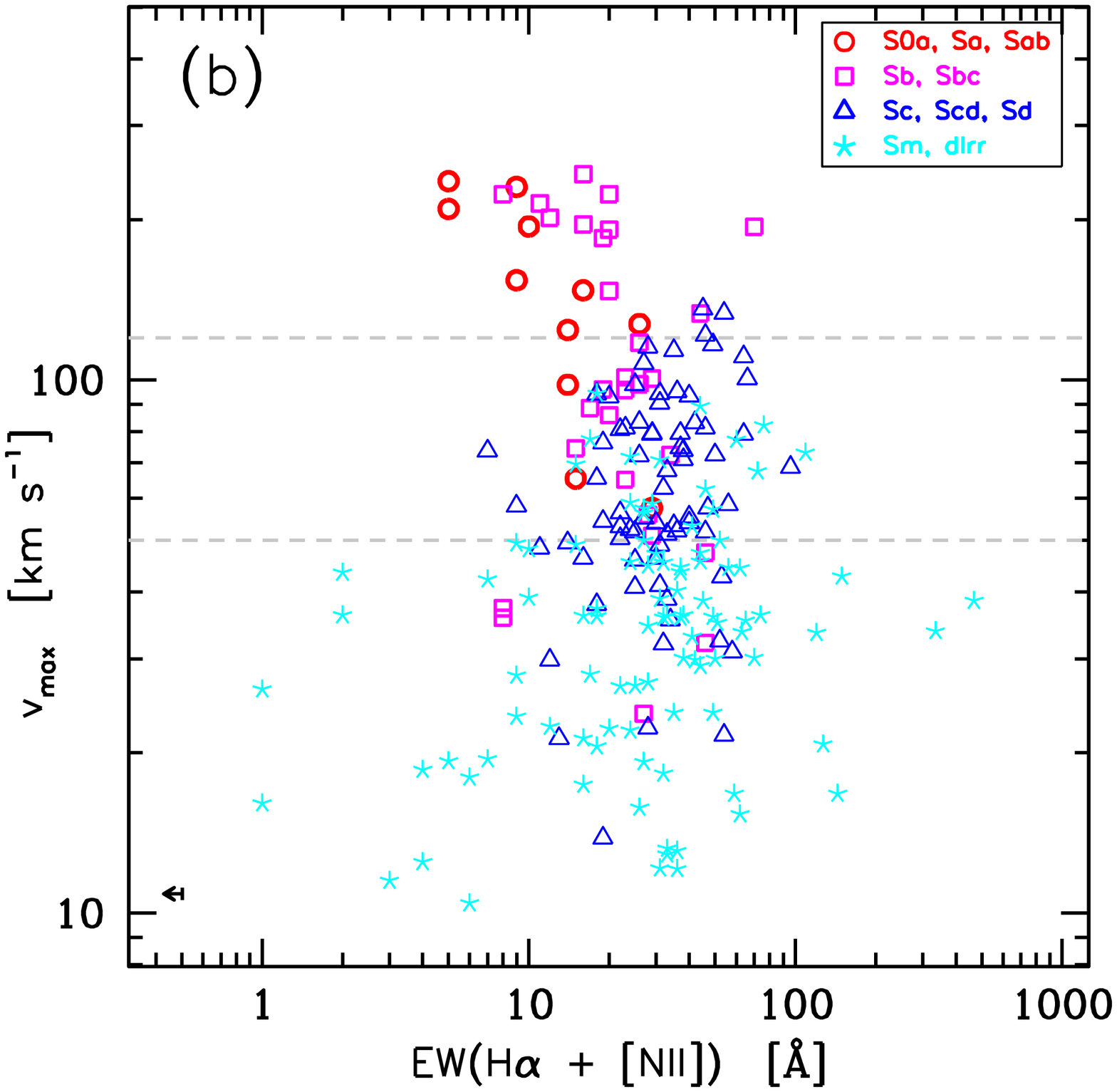}
\caption{The Local Volume star-forming galaxy sequence in the (a) $M_B$-EW and the (b) rotational velocity-EW planes.  Galaxies in the core sample of 11HUGS (i.e. those with T$\geq$0, D$<$11 Mpc and $|b|>20$\degr) are shown.  Gray dashed lines are drawn at $M_B = -19$ and $V_{max}$=120 \kms, and $M_B =-15$ and $V_{max}$=50 \kms to indicate the two transition regions discussed in the text.}
\end{figure*}
\section{Data}
To visualize trends in star formation across the 11HUGS sample, we 
examine the relationship of the H$\alpha$ EW with two indicators 
of the mass, $V_{max}$, the maximum circular rotational velocity, and $M_B$.   
We primarily use $M_B$ since $B$-band photometry is widely available in the literature 
and can be approximately reduced to a common (RC3) system for all of the 
galaxies in the sample. 
A well-known drawback with using the $B$-band light to trace 
stellar mass is that the mass-to-light ratio in the blue has a strong 
systematic dependence on the recent star formation history itself (e.g. 
Bell \& deJong 2001).  Examining the light in redder passbands, 
particularly in the near- and mid-IR, mitigates against these effects, 
and the 3.6 $\mu$m and 4.5 $\mu$m imaging being collected by the 
Local Volume Legacy will especially useful
in this regard.  In the meantime, we have also looked at 
the variation of the EW with $V_{max}$ as a 
qualitative consistency check.  $V_{max}$ 
traces mass gravitationally, and should in principle be independent of the recent star formation activity in individual galaxies.  The existence of the baryonic Tully-Fisher relation over the full mass-spectrum of late-type galaxies (e.g., McGaugh 2005) supports the use of $V_{max}$ as a mass indicator even in our dwarf dominated sample.  Using $V_{max}$ is not without its own weaknesses however.  One issue is that the maximal rotation velocity for low-mass systems is somewhat ill-defined since dwarfs are generally known to be solid body rotators (e.g. Skillman 1996).   Further, previous measurements of $V_{max}$ are limited, and are available for only $\sim$75\% of the sample.  Here we use $V_{max}$ values taken from the HyperLeda database, which have been extrapolated from single-dish 21-cm line widths (Paturel et al. 2003) using galaxies which have both line widths and rotation curves in the literature. 

To track the normalized star formation activity, we use the H$\alpha$ EWs from 11HUGS.  Recall that Balmer emission results from the recombination of nebular gas which has been photoionized by young, short-lived O and B stars.  Thus, the EW, which is given by the H$\alpha$ flux divided by the red continuum flux density, is an indicator of the strength of the current SFR relative to the total mass of stars (i.e., the specific SFR), and of a closely related quantity, the ratio of the current SFR to the past average SFR (i.e., the stellar birthrate parameter $b$).  The 11HUGS EWs used here have all been measured from narrowband imaging and are integrated over the entire extent of the galaxies so they are representative of global, galaxy-wide averaged values.  They do not suffer from the aperture effects that plague fiber-based spectral measurements.  

We provide details on the 11HUGS sample selection, observations, data processing, calibrations, measurements and completeness properties in Lee (2006), Kennicutt et al. (2008) and Lee \& Kennicutt (2008).

\section{Results}

The distribution of 11HUGS galaxies in the $M_B$-EW and $V_{max}$-EW planes are shown in Fig. 1.  $M_B$ is based on distances calculated from flow-corrected velocities assuming H$_{\circ}$=75 \kms Mpc$^{-1}$ when no direct distance estimates are available.  
No internal extinction corrections have been made to $M_B$.  We plot only those $\sim$300 galaxies in the sample that are (i) on the late Hubble sequence beginning with the S0a morphological type (T$\geq$0) and (ii) outside the Galactic plane ($|b|>20\degr$), and (iii) within 11 Mpc.  It is within these limits that we have tried to be as complete as possible in our compilation of currently known galaxies.  This core sample of 11HUGS is statistically complete down to B=15 (Lee 2006, Lee \& Kennicutt 2008).  
Although no corrections for internal extinction and the [NII] contribution to the EW have been applied to the data shown here, we have checked that such corrections (as described in Lee 2006 and Kennicutt et al. 2008) do not affect the results that follow.  
We have used different symbols to represent different morphological types (as given in the RC3) as indicated in the plot.  
Upper-limit symbols represent galaxies undetected in H$\alpha$. 
In Fig. 1b, galaxies that are nearly face-on ($i<30\degr$) are excluded because large inclination corrections result in highly uncertain $V_{max}$ values.

Over the full range of luminosities and circular velocities, star forming galaxies trace a continuous sequence in morphological type and EW with two characteristic transitions.  The trends are qualitatively consistent in both diagrams which gives us confidence that the features are real and not merely due to covariance between $M_B$ and the EW, and/or biases in the available $V_{max}$ data or its measurement.
   
Dwarf galaxies, mostly Magellanic spirals and irregulars, appear as the broad swath of light blue symbols at the bottom of the diagram.  They span the largest range of EWs observed in galaxies, and include both high EW starbursting blue compact dwarfs, as well as low EW systems that are barely forming stars and thought to be in an intermediate phase between the the gas-rich dwarf irregulars and gas-poor dwarf spheroidals.  In contrast, above $\sim50$ \kms and $M_B\sim-15$, a transition appears to occur where the EW distribution tightens.  This change is accompanied by a shift in the dominance of irregular morphologies to late-type spirals in the sequence.  Galaxies in the ``waist'' of the sequence have an average EW of $\sim$30 \AA.  Moreover, galaxies in which star formation has turned off or which are undergoing extreme starbursts are virtually absent.  

Finally, there appears to be another transition at $\sim120$ \kms and $M_B\sim-19$.  Above this velocity, the plume of points turns off towards lower EWs, and the range and dispersion of EW exhibited by galaxies increases again.  The sequence is then primarily occupied by bulge-dominated early-type spirals.

We compute the changing dispersions and mean EW values along the sequence by collapsing Fig. 1a in four coarse bins of $M_B$ chosen to correspond to the regimes defined by the two characteristic transitions (Table 1).  Measurements of both $M_B$ and EW are available for essentially all galaxies within the 11HUGS complete sub-sample, so the density of points in Fig. 1a is statistically representative of the Local Volume, down to our completeness limit of $-14.7.$  
\begin{table}[h]
\caption{Logarithmic H$\alpha$ EW Distribution Statistics}
\footnotesize
\begin{tabular}{crccccc}
\hline\\[-1ex]
&&\multicolumn{2}{c}{\it EW(H$\alpha$+[NII])}&&\multicolumn{2}{c}{\it EW(H$\alpha$)}\\
&&\multicolumn{2}{c}{\it no extinction correction}&&\multicolumn{2}{c}{\it extinction corrected}\\

\cline{3-4}\cline{6-7}
\\[-1ex]
\raisebox{0.ex}[0pt]{$M_B$}&\raisebox{0.ex}[0pt]{N}& $<$lg(EW)$>$&$\sigma$[lg(EW)]& &$<$lg(EW)$>$&$\sigma$[lg(EW)]\\
\hline
$[-22.0, -19.0$) & 27  & 1.17 & 0.40 & &1.15 & 0.43\\
$[-19.0, -17.0$) & 53  & 1.52 & 0.16 & &1.48 & 0.17\\
$[-17.0, -14.7$) & 87  & 1.50 & 0.17 & &1.49 & 0.20\\
$[-14.7, -13.0$) & 102 & 1.40 & 0.33 & &1.39 & 0.35\\
\hline
\end{tabular}
\end{table}

In all regimes, we find that the EWs are well characterized by log-normal distributions.  For the two intermediate luminosity bins which include galaxies with $-19\la M_B \la -15$, the means and dispersions of the 
logarithmic EWs are essentially the same, with values of $\sim$30\AA, and 1- and 3-$\sigma$ ranges 
(based on Gaussian fits to the distributions)
from about 20--50\AA, and 10--100\AA\ respectively. 
Above the turn-off, there is a drop in the mean EW by a factor of $\sim$2 to 15\AA, and an increase in the dispersion by a factor of 1.5.  At the opposite end of the sequence, 
there is a factor of two increase in the dispersion in the lowest luminosity bin.  There also appears to be a $\sim$25\% drop in the mean EW to 25\AA\ for these extreme dwarfs.  We have checked that although the fractional errors in the EW do become larger with both decreasing EW and decreasing luminosity, changes in the average random uncertainty do not drive the variation of the dispersion along the sequence. 
The average measurement error changes by several percent at most between luminosity bins, and this can cause an increase in the dispersion of only a few hundredths dex.  

\section{Discussion}

The plots in Fig. 1 provide a concise synthesis of the systematic 
dependencies of star formation activity with mass and morphology in the local universe.   
We comment on how these features reflect established empirical relationships, and 
re-frame open issues regarding star formation in sub-$L^*$ galaxies.

First, we consider the galaxies that flank the transition at
$M_B\sim-19$ and $V_{max} \sim 120$ \kms.  Changes 
in the mean star formation history as a 
function of mass and morphological type has been well established for
such systems (e.g., Kennicutt et al. 1994; Gavazzi
\& Scodeggio 1996), and the systematic drop in EW
above the transition indicates that star formation activity
is slowing down in galaxies above a certain characteristic
mass.  However, the abruptness of the transition to the 
``waist" of the sequence near $M_B = -19$ is somewhat 
surprising.
This is not an artifact of our small 11 Mpc sample;
the feature persists when we examine 
H$\alpha$ EW datasets which probe bright galaxies over larger volumes (e.g. James et al. 2004, Moustakas \& Kennicutt 2006).  

The narrowness of the sequence between $-19 \lesssim M_B \lesssim-15$ (50 \kms $\la$ \vmax 
$\la$ 120 \kms) places constraints on the range of normal variation in star formation activity as well as on the prevalence of
starbursts. 
It implies both a high degree of
temporal self-regulation within individual galaxies, and a small
dispersion in evolutionary properties among the galaxies. We explore these issues in detail in Lee (2006) and Lee \& Kennicutt (2008), but note here that based on the models of Kennicutt et al. (1994), the EW statistics imply that normal late-type spirals have $b$ between 0.1 and 2 with an average value of $\sim0.5$.  That is, intermediate luminosity disks commonly show factors of two to three fluctuations in their SFRs 
and are typically forming stars at half of their past average rate.  In the currently fashionable downsizing paradigm, these are the galaxies with sufficiently low mass and high gas content that have not yet exhausted their interstellar gas supplies, 
and have not yet merged to form more massive systems.  However such a scenario does not provide a ready explanation
for why the star formation histories of such galaxies should be
so homogeneous. 

Our results fit in well with recent work that has
also aimed to characterize the star-formation properties of local galaxies. 
Brinchmann et al. (2004) used 
Sloan Digital Sky Survey (SDSS) spectra for $\sim10^5$
galaxies and extensive modeling to compute H$\alpha$-based SFRs.  They show that $b$ is nearly constant for $10^{8} \lesssim M_{\ast} \lesssim 10^{10}$, but that the distribution broadens and shifts to lower values beginning at 
$M_{\ast} > 10^{10}$ \msun.  Salim et al. (2007) combined
GALEX UV and SDSS optical photometry for $\sim$50,000 
galaxies and found that the UV-based specific SFRs also occupy a narrow sequence with mass and begin to decline on average in the same regime. 
Noeske et al. (2007) show that the population of galaxies
with a narrow dispersion in SFRs at a given M$_{\ast}$ persists back to redshifts $z \sim 1$,
and they refer to this feature as the ``main sequence'' 
of star-forming galaxies.  Our new H$\alpha$ narrowband imaging 
observations provide complementary confirmation of these results -- they demonstrate that the trends also correlate with morphology and are already evident in the observable tracers of the physical quantities.  

Work on understanding the physical origin of the transition at $M_B\sim-19$ and $V_{max} \sim 120$ \kms\  (i.e., $M_{\ast} \sim 10^{10}$ \msun) is on-going
and generally involves investigating gas inflow and/or energy feedback from supernovae and AGN into the ISM (e.g., Keres et al. 2005, Croton et al. 2006).  Processes associated with bulge formation are also likely important, as we observe that the transition is accompanied by a shift from late- to early-type disks.
We also note the intriguing coincidence 
of other disk properties which show 
transitions near  $M_B\sim$-19 and $V_{max}\sim$120 km s$^{-1}$.
The mass-metallicity relation undergoes a change in slope 
at approximately the same luminosity (Garnett 2002, Tremonti et al. 2004).
Edge-on disk galaxies show distinct
structures above and below $V_{circ} \sim 120$ \kms, with
more massive galaxies showing thin stellar disks
and well-defined thin dust lanes, while more slowly rotating
disks 
exhibit thick stellar disks and no
dust lanes (Dalcanton et al. 2004).
This velocity has also been identified as the critical scale
that subdivides systems that can retain the ejecta
from supernova feedback ($V_{circ} > 120 - 130$ \kms), 
as opposed to those that may lose some
of their metals into the 
IGM (e.g., Martin 1999, Dekel \& Woo 2003).  

Below $M_B\sim$-15 and $V_{max}\sim$50 km s$^{-1}$ the
dispersion 
increases again.   With smaller representative samples Kennicutt et al.
(1994) and Hunter (1997) had already noted that dwarf irregulars show a large scatter in their relative star formation properties as compared with 
spiral galaxies.
In contrast to the upper mass
transition however, this feature has not been as well studied, 
because large surveys such as SDSS do
not adequately sample the faint end of the galactic luminosity 
function, and previous studies of SFRs of dwarf galaxies lacked
the completeness needed for statistical analysis.  
The lower transition is nearly as pronounced as the 
upper transition, and 
it is interesting to ask whether it is indicative
of another change in the dominant physical processes that 
regulate star formation.  If so, then the 
transitions mark the boundaries 
of three regimes where distinct modes of star formation operate.

A key question however is whether the increase in scatter
at low masses reflects an increase in statistical
fluctuations in massive star formation for galaxies with SFRs
below $\sim$0.01 \msun~yr$^{-1}$, rather than a fundamental physical 
change in the degree of regulation  in the global SFR.
One possibility is that the increased scatter may
arise from a breakdown in the reliability of 
H$\alpha$ emission as a quantitative SFR measure when 
star formation is so low that it does not fully populate
the upper end of the IMF, as suggested by Weidner
\& Kroupa (2006) and Pflamm-Altenberg et al. (2007).
However such effects are unlikely to account
for the transition we observe.  At $M_B = -15$ the average
SFR of the galaxies in our sample is approximately 0.03
\msun~yr$^{-1}$.
A galaxy with the mean SFR and a Salpeter IMF would contain
approximately 300 30~\msun\ main sequence stars at any one
time; if these stars formed 
independently, statistical fluctuations in their
SFR of $<$10\% would be expected, negligible in comparison to the
observed dispersion in SFRs at any galaxy mass.  
If instead stochastic effects are dominated by the 
clustering of star formation,
it is the $\sqrt{N}$ statistics of
the number of clusters or HII regions that are relevant.
However at the transition luminosity the average galaxy
in our sample possesses at least several HII regions,
and often tens or more.  Thus, this effect
alone would only produce a dispersion of a few
tens of percent, again too small to account for the 
hundredfold range of SFRs observed.  
Still such stochasticity is important for the faintest
galaxies in our sample ($M_B\gtrsim-13.5$), with SFRs $\lesssim$0.001~\msun,
and we investigate this further in Tremonti et al. (2008).

If instead the increased variability of the SFR at
low galaxy masses is physical, then what mechanisms are responsible?  
The answer to this question is still unclear although there has
been much work on star formation in gas-rich dwarfs over the past three decades (e.g., Hunter 1997, Hunter \& Elmegreen 2004, and references therein).  The problem is particularly vexing because of the lack of an easily identifiable "second parameter" that correlates with the increased scatter; i.e., dwarf irregulars with very similar global properties (e.g., environment, gas content, structural properties) can have a wide range in normalized star formation activity
(but see Papaderos et al. 1996, van Zee 2001).  Our preliminary examination of EW with proximity to known neighbors, gas fraction and the concentration of H$\alpha$ emission has also not uncovered any obvious trends.  It is reasonable to speculate, however, that spiral structure plays a key role in regulatory processes since the increase in scatter at the low end roughly tracks a shift in morphology from spiral to irregular types.

The general absence of a relationship of activity in dwarf irregulars with environment outside rich groups and clusters (e.g., Lee et al. 1999, Hunter \& Elmegreen 2004) also suggests that internal processes may be responsible for the increase in scatter.  One possibility is that feedback from massive stars (e.g. stellar winds, supernova shocks) 
has a greater negative impact on the ISM of 
physically small, kpc-sized systems, with the result that star 
formation is burstier in dwarf galaxies than in larger spirals.  
Gerola et al. (1980) originally explored such issues in their 
models of stochastic self-propagating star formation, which interestingly show a characteristic stabilizing transition in fluctuations in the mean star formation rate for galaxies with $M_B\ga$-15.  
This result, however, is 
dependent on the assumed 
``cell" size and the refractory period (the time during 
which negative feedback inhibits the formation of new stars), which 
are free parameters in their toy models.  The models of Mac Low \& Ferrara (1999)
incorporate more realistic prescriptions for the 
density distributions of the gas, stellar and dark matter in dwarf 
galaxies.  Again, they find that below a rotational velocity 
of $\sim$30 km/s, feedback disrupts and blows out the ISM from the 
galaxy.  Most recently, Stinson et al. (2007) have modeled
the effects of supernova feedback in haloes with virial radius velocities from 10-30 km s$^{-1}$ using SPH + N-body simulations.  These models show there is an increase in episodic behavior with decreasing 
halo mass.

This line of work offers a promising explanation for the low-mass broadening of the EW distribution, but a more careful comparison 
of model outputs and observations are needed.  In particular, the combination of 
star formation histories derived from
observations of the resolved stellar populations of the nearest
galaxies (e.g., from the ACS Nearby Galaxy Survey Treasury, ANGST, Dalcanton 2006) and the distribution of global properties from 11HUGS 
provides complementary constraints. Viable models 
will simultaneously need to reproduce features of the temporally and spatially resolved star formation histories from ANGST, the trends in the relative SFR with mass as reported here,
as well as the constraints on the starburst duty cycle which will be presented in upcoming papers.

\acknowledgments
It is a pleasure to thank S. Salim, D. Zaritsky, L. van Zee and J. Salzer
for feedback on early versions of this paper.  
We also thank the referee J. Brinchmann for a careful reading of the manuscript and useful comments.  Financial support from NSF grants AST-9617826 and AST-0307386 is acknowledged.

\end{document}